# Semiparametric Spatiotemporal Model with Mixed Frequencies

Vladimir A. Malabanan

School of Statistics, University of the Philippines Diliman, Quezon City, Philippines

Joseph Ryan G. Lansangan

School of Statistics, University of the Philippines Diliman, Quezon City, Philippines

Erniel B. Barrios
ORCID:0000-0003-4209-2858

School of Statistics, University of the Philippines Diliman, Quezon City, Philippines
corresponding author: ebbarrios@up.edu.ph

Abstract

In modelling time series data coming from different sources, frequencies can easily vary since some variable can be measured at higher frequencies, others, at lower frequencies. Given data measured over spatial units and at varying frequencies, we postulated a semiparametric spatiotemporal model. This optimizes utilization of information from variables measured at higher frequency by estimating its nonparametric effect on the response through the backfitting algorithm in and additive modelling framework. Simulation studies support the optimality of the model over simple generalized additive model with aggregation of high frequency predictors to match the dependent variable measured at lower frequency. With quarterly corn production and the dependent variable, the model is fitted with predictors coming from remotely-sensed data (vegetation and precipitation indices), predictive ability is better compared to two benchmark generalized additive models.

**Subject Codes:** 62G08 37M05 37M10 68W01

**Acknowledgement:** Simulation done in R were impleted in CoARE Facility of the DOST-Advanced Science and Technology Institute (DOST-ASTI) and the Computing and Archiving Research Environment (CoARE) Project.

# 1. Introduction

It is projected that the world population will reach 8.5 billion by 2030 (UN Report, 2015). This presents challenges in achieving sustainable development, requiring more resources to support the growing needs including food security. Achieving food security requires adequate food availability, access and use, and agriculture plays a key role in this. Increased food production can be achieved by raising crop yields on existing farm land, expanding crop production area, or both (Wart et al., 2013). But achieving such crop production increases will not be as easier as it is in the past. In fact, though historical evidence suggests that the growth of global agricultural production has so far been more than sufficient to meet the growth of demand, it should not be taken for granted the threat of uncertainties from growing resource scarcity and climate change, (Alexandratos and Bruinsma, 2012; World Bank, 2008). Thus, reliable estimates of agricultural production are important to aid policy makers in situational market analysis and decision-making to ensure that food supply meets the demand for consumption.

Food and Agriculture Organization of the United Nations (FAO-UN) identifies main sources of data on agricultural production which include censuses of agriculture and livestock, crop estimation surveys, farm management and cost of cultivation studies, household surveys and various returns collected by administrative agencies concerned with prices and production relating to agriculture. However, traditional way of generating estimates for agricultural production can be challenging and costly, and possible difficulty in obtaining accurate data. Craig and Atkinson (2013) made a study of crop area estimation procedures that are used by national statistical agencies across different countries. They analyzed several methods of crop production estimation from voluntary crop reports to statistically designed surveys and census to advanced approaches such as the use of remote-sensing and GIS data. Schedule of data collection has been observed to be a major challenge in estimating agricultural production. Area estimation can pose problems to developing countries which have limited resources to conduct surveys and censuses. In the Philippines, rice and corn production estimates are based on Rice and Corn Production Survey (RCPS) which is conducted quarterly, production data are collected on the month following the reference quarter. This means that crop yield information will only be reported at least a month after the reference period.

Given these challenges in current methods in collecting agricultural statistics, FAO-UN has acknowledged the need of modernization of agricultural statistics by "*moving from a paradigm of producing best estimates possible from a survey to that of producing the best possible estimates to meet user needs from multiple data sources.*" In the International Conference on Agricultural Statistics held in Rome, Italy in 2016, FAO-UN emphasized importance of combining different methods (and sources) aside from survey/census and leveraging on the use of technology. These other methodologies noted (but not limited to) are mathematical models, big data and remote sensing.

*Spatiotemporal Modeling of Agricultural Production*

Agricultural production is a complex system that needs to consider both space and time elements. Production of adjacent farm lands can be correlated due to similarities on their weather and environmental conditions including exposure to pests, and interplay to affect growth of agricultural crops. Furthermore, nearby areas may share same similar endowment of resources critical to agricultural production, affecting each other's yield. There is also a temporal relationship due to crop rotation and seasonal variation of weather. Simple estimation procedures which assume independence of observations may not aptly capture the dynamics of spatial and temporal associations of exhibited in yield of crops.

Spatiotemporal modeling in the context of crop yield and production estimation has been thoroughly studied in the past. Landagan and Barrios (2007) postulated a spatiotemporal model estimated by embedding the Cochran-Orcutt procedure in a backfitting algorithm. The proposed procedure was found to give superior performance in estimating rice production. Similar procedure was extended by Dumanjug (2007) this time, estimating corn yield. Other techniques used for spatiotemporal estimation of agricultural data are fuzzy multicriteria model (Guan et al., 2016) and bayesian models (Ozaki et al., 2008).

The use of remote sensing in the estimation of agricultural yield has also been studied and explored since satellite data have become available and more accessible. National Agricultural Statistics Service (NASS) in the United States has identified three critical components in using remote sensing in estimation of agricultural statistics. First is the ground truth that is critical in the validation of approximates. Second is the identification of source of satellite imagery with factors such as frequency and spatial and temporal dimensions as well as cost of data collection should be considered. Lastly, analysis of these data may need investment in technology as extraction, parsing, and pre-processing techniques may be computationally expensive.

Vegetation indices extracted from satellite images provides good indication of potential crop yield (Carfagna and Gallego, 2005; Johnson, 2013; Huang et al., 2013). Statistical agencies are also exploring to adopt remote sensing data in reporting of official agricultural statistics. In Malawi, a methodology for incorporating the use of remote sensing and satellite imagery in its programs for crop production estimation to improve accuracy of crop production statistics has been piloted, noted (Mwanaleza, 2016).

Given the high temporal resolution of remote-sensed data available, there are many approaches in using vegetation indices from satellite data in estimating agricultural production. High frequency data parsed from satellite images are typically summarize into indicators that follow the same temporal frequency as the response variable. This method however, is prone to losing vital information on crop yield contained in remotely-sensed data. In fact, it will be important to see vegetation indices at a more frequent rate so that it can help explain each phonological stage of crop growth (Dela Torre and Perez, 2014).

Given data from different sources and with varying frequencies, the dependent variable has lower frequencies than the predictors, we postulate a spatiotemporal model. Predictors with same frequency as the dependent variable are incorporated into the model with the parametric structure, while the predictors with higher frequencies are postulated in a nonparametric form. The semiparametric model is then estimated through a hybrid method in the framework of the backfitting algorithm. This modelling and estimation strategy aims to avoid unnecessary aggregation of high frequency data to complement those of the low frequency data. This will optimize the utilization of information in the construction of a model in a predictive analytics environment. The model is used in estimating crop yield with high frequency remotely-sensed data as predictors.

## 2. Some Modeling Strategies

### 2.1   The Backfitting Algorithm

Hastie and Tibshirani (1986) introduced a class of generalized additive models that deviates away from likelihood-based estimation in modeling the response $Y$ as a linear function of a set of covariates $X_1, \ldots X_p$. Dependencies of $Y$ on $X_1, \ldots X_p$ is estimated non-parametrically is an additive model, i.e., in lieu of the linear form, a sum of smooth functions $\sum_1^p s_j(X_j)$ is used. Similarly, even a linear regression model can be estimated as an additive model in the form:

$$E(Y|X_1, X_2, \ldots X_p) = s_o + \sum_1^p s_j(X_j) \tag{1}$$

where $E\left(s_j(X_j)\right) = 0$ for every j. Hastie and Tibshirani (1986) also proposed the backfitting algorithm, an iterative process of estimation that took advantage of the additive structure of the model in Equation 1. The method is based on the algorithm introduced by Breiman and Friedman (1985) where individual smoothing functions are fit separately against the response, and done iteratively until convergence, or when the smoothing functions no longer register changes in subsequent iterations.

The process as illustrated by Hastie and Tibshirani (1986) assumes that the additive model is correct and that $s_o, s_1(\cdot), \ldots, s_{j-1}(\cdot), s_{j+1}(\cdot), \ldots, s_p(\cdot)$ are known and can be estimated given n observations. Partial residuals are computed in every iteration following Equation 2.

$$R_j = Y - s_o - \sum_{j \neq k} s_k(X_k) \qquad (2)$$

where $E(R_j|X_j) = s_j(X_j)$ and minimizes $E(Y - s_o - \sum_{k=1}^{p} s_k(X_k))^2$. Specific details of the algorithm is given as follows:

a. Initialize: Let $s_o = E(Y), s_1^1(\cdot) \equiv s_2^1(\cdot) \ldots \equiv s_p^1(\cdot) \equiv 0, m = 0$.

b. Iterate: $m = m + 1$

For $j = 1$ to $p$ do:

$$R_j = Y - s_o - \sum_{k=1}^{j-1} s_k^m(X_k) - \sum_{k=j+1}^{p} s_k^{n-1}(X_k)$$

$$s_j^m(X_j) = E(R_j|X_j)$$

c. Converge: Iterate until $RSS = E(Y - s_o - \sum_{j=1}^{p} s_j^m(X_j))^2$ fails to decrease.

Convergence and consistency of the backfitting algorithm in an additive model with a general linear smoother has been proven by Buja et al. (1989). Note further that the backfitting algorithm will work in the time-series context provided that the dependence structure is not quite strong (Chen and Tsay ,1993). Asymptotic properties of backfitting estimators are also studied and shown to achieve the same rate of convergence as those of univariate local polynomial regression (Opsomer, 2000)

## 2.2  *Semiparametric Modeling*

Parametric models assume a definite structure defined by a finite set of parameters, $\theta$. Given the parameters, future characterization of $y$ are made independent of the observed data, $\mathcal{D}$ used in estimating these parameters, i.e.,

$$P(y|\theta, \mathcal{D}) = P(y|\theta) \qquad (3)$$

The parameter $\theta$ capture everything there is to know about the data $\mathcal{D}$. Knowing the function that describes the relationship between the response and explanatory variables is then crucial in parametric estimation. In compiling data from various sources, this functional relationship is sometimes difficult to assume specially when the data is large and is often confounded with heterogeneity. This necessitate to consider more flexible structure in nonparametric models, where data is used to infer an unknown function while making as few assumptions as possible. This usually means using statistical models that are infinite-dimensional (Wasserman, 2006). Nonparametric regression estimators are very flexible but precision decreases greatly with increasing number of explanatory variables in the model (Hardle et al, 2004). To overcome this, features of parametric and nonparametric techniques are combined into a semiparametric model.

A semiparametric model is a hybrid of the parametric and nonparametric structure in the construction of a statistical model. It distinguishes between the "parameters of interest", which are finite-dimensional, and infinite-dimensional "nuisance parameters", which are treated nonparametrically. Similar to the parametric and nonparametric approaches, semiparametric modelling shares the similar advantages and disadvantages of each, i.e., estimators of the parameters of interest for semiparametric models are consistent under a broader range of conditions than for parametric models but generally less efficient than maximum likelihood estimators for a correctly-specified parametric model (Powell, 1994). Newey et al. (1990) used semiparametric methods in analyzing several datasets to study the empirical ability of the method, noted that estimates based upon semiparametric restrictions were comparable to their parametric counterparts, but with larger standard errors.

## 2.3  *Modeling Time Series Data with Mixed Frequency*

There is a rich literature on how to estimate a time series model with mixed or varying frequencies among the dependent and independent variables. Foroni and Marcelino (2013) provided a survey of methods used for mixed frequency data in the context of econometrics. The most commonly used technique include bridge equations and MIxed DAta Sampling (MIDAS), mixed frequency VAR and mixed frequency factor models. They further noted that though most usual solution to ensure that data are of the same frequency is through aggregation, a lot of potentially useful information can be lost, and possibly trigger misspecification error into the model. Even though there is a concensus that exploiting data at varying frequency matters, there is no clear method that is universally superior.

These methods included in the review of Foroni and Marcelino (2013) emphasized notable performance in empirical applications. Kuzin et al. (2011) compared mixed-data sampling (MIDAS) and mixed frequency VAR (MF-VAR) in nowcasting and forecasting quarterly GDP growth in the Euro area using monthly indicators. MIDAS tends to give better short-term forecasts while MF-VAR performs better on longer horizons. Clements and Galvao (2012) used MIDAS to forecast US output growth using monthly predictors.

## 2.4   *Spatiotemporal Models*

Spatiotemporal models arise when data are collected across time as well as in space. Landagan and Barrios (2007) proposed a hybrid estimation procedure for spatiotemporal model via the backfitting algorithm. The idea is to alternately estimate the parameter effect of geophysical covariates, spatial variables and the temporal dependence structure. The method was illustrated in predicting corn yield in the Philippines. The postulated model give better forecast than some benchmark models. This was extended by Bastero and Barrios (2011) to take into account structural change in a spatiotemporal system of disease incidence and Villejo et al (2017) to consider dynamic spatiotemporal models with structural change in agricultural production.

Sahu and Mardia (2005) provided a survey of the latest trends in modeling spatiotemporal data, noted that techniques that were largely developed includes hierarchical Bayesian approaches for environmental pollution studies, spatiotemporal autoregressive moving average (STARMA) for climate and other environmental applications, spatially varying linear model of coregionalization (SVLMC) for sociological and socioeconomic studies and spatiotemporal modeling with drift for geostatistics and hydrology.

## 3. Spatiotemporal Model with Mixed Frequency

Consider the spatiotemporal model proposed by Landagan and Barrios (2007) given as follows:

$$y_{it} = \beta Z_{it} + \gamma W_{it} + \varepsilon_{it} \qquad (4)$$

where $Y_{it}$ is the response variable from location $i$ at time $t$, $Z_{it}$ the set of covariates from location $i$ at time $t$, $W_{it}$ the set of variables in the neighborhood system of location $i$ at time $t$, and $\varepsilon_{it}$ the error component that would take the disturbances. The error components are further assumed to exhibit temporal dependency in an autoregressive behavior as $\varepsilon_{it} = \rho \varepsilon_{it-1} + a_{it},\, , |\rho| < 1 \quad a_{it} \sim IID(0, \sigma_a^2)$.

Consider a dataset where at least of the predictors are measured at higher frequency than the dependent variable. Model (4) is modified to accommodate varying frequency of measurement of the response and different covariates. Since one covariate is measured at a higher frequency than the response, this is postulated to affect the dependent variable in a nonparametric component. Without aggregation of higher frequency predictor, this preserves the information contained in the predictor so it can completely account whatever variability in the dependent variable it is capable of explaining. The parameter β and temporal (autoregressive) parameter ρ are assumed to be constant across the spatial units. The proposed semiparametric spatiotemporal mixed frequency (SemiparMF) model is given in Equation (5).

$$y_{it} = \sum_{k=1}^{K} f(X_{it_k}) + \beta Z_{it} + \gamma W_{it} + \varepsilon_{it} \qquad (5)$$

$$\varepsilon_{it} = \rho \varepsilon_{it-1} + a_{it}, |\rho| < 1 \quad a_{it} \sim IID(0, \sigma_a^2), i = 1,\ldots,n, t = 1,\ldots T, k = 1,\ldots,K$$

where

$Y_{it}$ = response in spatial unit $i$ at time $t$

$X_{it_k}$ = covariate measured at higher frequency of spatial unit $i$ at subperiod $k$ of time $t$, $k=1,2,...,K$.

$f(\cdot)$ = a continuous function in $X_{it_k}$

$Z_{it}$ = covariate measured in same frequency as response in spatial unit $i$ at time $t$

$W_{it}$ = neighborhood system where spatial unit $i$ belongs at time $t$

$\varepsilon_{it}$ = error terms for spatial unit $i$ at time $t$

The following assumptions are further considered:

i. constant nonparametric component effect across units across time

ii. constant parametric component ($\beta$) effect across units and across time

iii. constant neighborhood variable ($\gamma$) effect across units and across time

iv. constant temporal effect ($\rho$) across units

The constant effect of both the nonparametric and parametric components of the model assumes that the effect of the covariates does not significantly change across space and over time. In predicting agricultural production, the effect of vegetation index data ($X_{it_k}$) with higher temporal resolution is assumed to be constant across provinces (spatial units) since these data are extracted from a single satellite image. This implies that the relationship of vegetation index for each phonological stage of corn may not vary across different provinces, (Dela Torre and Perez, 2014). The component $Z_{it}$ is measured at each spatial unit and is presumed to be constant across provinces and across time (example of this type of variable can be total area harvested of corn). This further implies that if area harvested is high $t=0$, it will be more likely to remain high at further time point $t = 1, 2, ...., T$, hence constant $\beta$ over time. $W_{it}$ is defined to be variable in the neighborhood system. In an agricultural setup, geographical proximity provides a suitable basis for the definition of a neighborhood. In corn yield example, this can be average quantity of inorganic fertilizer applied to crop in the region, its effect to crop production is assumed to not vary across time. Temporal effect is further assumed to be constant across provinces.

The SemiparMF model can also be generalized to account for stylized facts about the time series like seasonality by altering the specification of the temporal dependencies in the error term, say, seasonal autoregressive model.

The model in Equation (5) can be estimated through hybrid methods embedded into the backfitting algorithm. Nonparametric components (for high frequency predictors) are estimated first, followed by the covariate effect for each spatial unit ($\beta_i$) and the spatial parameter ($\gamma_i$) and finally, the temporal parameter $\rho_i$. Though covariate effect $\beta$, spatial parameter $\gamma$ and temporal parameter $\rho$ are estimated per spatial unit, monte carlo estimates are then obtained. This hybrid estimation algorithm is summarized as follows:

1. Given the response $\{y_{it}\}, i = 1,2,...n; \ t = 1,..T$ and $\{x_{itk}\}, i = 1,2,..n; \ t = 1,2,..T; \ k = 1,2,..K$, the covariate measured at high frequency, ignore all other components of the model to estimate the nonparametric component using smoothing splines. Compute the fitted values and the corresponding residuals from Equation (6).

$$e_{it}^1 = Y_{it} - \sum_{k=1}^{K} \widehat{f_{it_k}(X_{it})} \tag{6}$$

2. Estimate the covariate effect, $\beta$ and the neighborhood covariate effect $\gamma$ per spatial unit, i.e., fit a regression model using the computed residuals in the in Equation (6) as the response and $Z_{it}$ and $W_{it}$ as covariates using weighted least squares. Then compute the average and take it as estimates of the covariates ($\hat{\beta}$ and $\hat{\gamma}$). With the nonparametric component estimated in Step 1 and estimates $\hat{\beta}$ and $\hat{\gamma}$ in Step 2, compute the updated fitted value and the corresponding residuals from Equation (6). These residuals still contain information on the true error and the temporal parameter ($\rho$).

3. For each location, estimate the autoregressive model AR(p), e.g., $\rho_i$ if $p = 1$. Take $\hat{\rho}$ as average of these estimates.

4. After Step 3, we now have a smoothing function summarizing the effect of $X_{it_k}$ and the parameter estimates $\hat{\beta}, \hat{\gamma}$ and $\hat{\rho}$. We compute the final fitted value and the corresponding residuals. A new dependent variable will be computed by adjusting for the temporal component from $y_{it}^{new} = y_{it}^{orig} - \hat{\rho}_i e_{i,t-1}$, where the residual is initialized with 0.

5. The algorithm iterates from Step1 using the updated response computed in Step 4 (less the temporal effect). In updating the values of the residuals, use the original values of $Y_{it}$. In Step 4, the response variable will be updated using the original values of the response variable and the updated estimates of the error terms in Step 3. The iteration continues until there is minimal changes in the MSPE (<1%).

This iterative estimation procedure assumes that the covariate with higher frequency drives more variability in the response than the other covariates, explaining why it is estimated first. The information contributed by each independent variable is isolated from the dependent variable at each step of estimation. This procedure is repeated until convergence or when there are no more changes in the nonparametric function and parametric estimates of the model.

## 4. Simulation Studies

The SemiparMF model and estimation procedure are evaluated using simulated data when temporal resolution of the higher frequency variable is low ($K = 3$), such in the case of having quarterly response and monthly covariate; or high ($K = 12$) or when response is measured yearly and covariates are observed monthly. Response variable was generated following Equation (5), specifically,

$$y_{it} = a * \sum_{k=1}^{K} h(X_{it_k}) + b * Z_{it} + c * W_{it} + m * \varepsilon_{it} \qquad (7)$$

where $X_{it_k}$ is the higher temporal resolution covariate generated from U(0,1)

$Z_{it}$ is the covariate with same frequency as the response that is generated from $N(100,10)$

$W_{it}$ is neighborhood variable generated from $Po(\lambda)$, $\lambda = 50$

$h(\cdot)$ is the functional form of the simulated $X_{it_k}'s$ (eg, linear, quadratic, exponential)

a = weight of the function of high frequency covariate

b = weight of the covariate $Z_{it}$

c = weight of the neighborhood variable

m = used to introduce misspecification error, known to induce bias in parametric modeling

The weights of each covariate were set to have several scenarios that varies their contribution to the variability of the response. Response can be dominated by the covariate in higher frequency over the other covariates. It can also be the case when the variation of the response variable is caused significantly by the spatial covariate that is shared among groups of observations. Equal contribution of the covariates to the variability of the response is also simulated.

Since covariates cannot always be assumed to be free from temporal dependencies in a spatiotemporal setting, they are allowed to assume temporal correlations in the simulations. Specifically, the independent variables are simulated to be uncorrelated, has strong temporal correlation ($\rho = 0.8$), or has weak temporal correlation ($\rho = 0.1$). The functional form of the higher frequency covariate was also allowed to vary from linear, quadratic and exponential forms.

Furthermore, the behavior of the algorithm is considered for balanced (N=T) and unbalanced data sets (N>T, N<T).

The errors are generated via AR Sieve in two scenarios with $\rho = 0.5$ and $\rho = 0.9$. Each simulation scenario is replicated 100 times. Table 1 summarized the simulation boundaries used in the assessment of the SemiparMF model and estimation methodology.

The SemiparMF model is compared to ordinary general additive models in terms of mean squared prediction error (MSPE) and mean absolute percentage error (MAPE). Two additive models were simultaneously estimated to be compared with the hybrid method: (1) generalized additive model where covariates in higher frequency are individually estimated via spline smoothing while other covariates are estimated parametrically and estimation is done with one iteration only; (2) generalized additive model where covariates in higher frequency are summarized to their means first before estimating non parametrically while other covariates are estimated parametrically and estimation is done with one iteration only. Predicted values are calculated by simply adding up the scores in the estimated nonparametric function and the linear combination of the parameter estimates and their corresponding covariates.

Table 1. Simulation Settings

| Parameter | Parameter Settings | No. of Settings |
|---|---|---|
| No. of subtime points per unit time | a. K=3<br>b. K=12 | 2 |
| Contribution of each model component to the response (f(x)-z-w-e) | a. Equal contribution (30-30-30-10)<br>b. Dominating high frequency covariate (50-20-20-10)<br>c. Dominating other covariate (20-50-20-10)<br>d. Dominating neighborhood covariate (20-20-50-10) | 4 |
| Sample size and length of time series | a. Balanced data (N=50, T=50)<br>b. Longer time series than number of observations (N=30, T=50)<br>c. More observations than length of time series (N=50, T=30) | 3 |
| Correlation of the error terms | a. Moderate ($\rho = 0.5$)<br>b. Strong ($\rho = 0.9$) | 2 |
| Functional form of the $h(X_{it_k})$ | a. Linear $\sum(X_{itk})$<br>b. Quadratic $\sum(X_{itk}^2 + X_{itk})$<br>c. Exponential $\sum(e^{X_{itk}})$ | 3 |
| Misspecification Error | a. With misspecification (m=10)<br>b. Without misspecification (m=1) | 2 |
| Nature of covariate | a. No temporal autocorrelation<br>b. AR with strong correlation ($\rho = 0.8$)<br>c. AR with weak correlation ($\rho = 0.1$) | 3 |

## 5. Results and Discussion

There are a total of 864 different scenarios for various combination of settings in Table 1. The discussion is partitioned into different nature of the covariates: nonautocorrelated covariates; covariates with weak temporal correlation ($\rho = 0.1$), and; (3) covariates with strong temporal correlation ($\rho = 0.8$). In each partition, focus will be on the comparison of the scenarios depending on the frequency of occurrence of the more frequent covariate ($K = 3$ and $K = 12$) as well as the temporal correlation of the error terms. The performance of the algorithm is gauged in terms of its predictive ability as measured by MSPE and MAPE. It is then compared to ordinary generalized additive model (GAM – Ordinary) and another generalized additive model with the high frequency variable summarized to its mean (GAM – Summarized).

### *5.1 Covariates without autocorrelation*

Table 2 shows the average mean squared prediction error and mean absolute percentage error for all scenarios when there is no autocorrelation on the covariates and at various levels of autocorrelation of the error terms and levels of $K$. It is interesting to note that when the frequency of the other higher temporal resolution covariate is low e.g., $K = 3$, we can opt to choose the ordinary generalized additive model with summarized higher frequency covariate as it gives lower MAPE compared to the other two models. This is particularly true when the temporal correlation of the error terms is weak ($\rho = 0.5$) and functional form of more frequent covariate is either linear or exponential. However, note that when there is strong autocorrelation in the series, the SemiparMF model estimated with the backfitting algorithm yield the lowest average MSPE and MAPE. The SemiparMF model further shown better performance when $K = 12$ over the two generalized additive models. It has superior predictive ability even with misspecification errors.

Table 2. MSPE and MAPE for Uncorrelated Covariates, Varying Functional Forms of High Frequency Covariates

| | Error Correlation | Functional Form | No. of Subtime Points | SemiparMF | | GAM - Ordinary | | GAM - Summarized | |
|---|---|---|---|---|---|---|---|---|---|
| | | | | MSPE | MAPE | MSPE | MAPE | MSPE | MAPE |
| Without Misspecification Error | $\rho=0.5$ | Linear | 3 | 586.72 | 12.0% | 685.92 | 13.1% | 314.9 | 8.8% |
| | | | 12 | 210.15 | 7.5% | 365.60 | 10.0% | 320.0 | 9.3% |
| | | Quadratic | 3 | 210.69 | 7.6% | 1036.66 | 15.3% | 325.0 | 8.6% |
| | | | 12 | 171.50 | 6.1% | 397.31 | 10.2% | 325.7 | 9.2% |
| | | Exponential | 3 | 962.09 | 14.6% | 368.63 | 10.1% | 319.9 | 9.4% |
| | | | 12 | 249.24 | 8.0% | 346.19 | 8.8% | 335.4 | 8.6% |
| | $\rho=0.9$ | Linear | 3 | 1002.00 | 10.9% | 4795.83 | 26.2% | 4437.3 | 25.1% |
| | | | 12 | 370.41 | 6.4% | 4412.09 | 26.1% | 4441.6 | 26.3% |
| | | Quadratic | 3 | 370.29 | 6.4% | 5147.29 | 26.3% | 4448.3 | 24.3% |
| | | | 12 | 308.07 | 5.2% | 4443.58 | 25.7% | 4447.4 | 25.8% |
| | | Exponential | 3 | 1602.76 | 13.5% | 4478.69 | 26.5% | 4442.7 | 26.3% |
| | | | 12 | 433.75 | 7.0% | 4393.23 | 24.0% | 4457.4 | 24.2% |
| With Misspecification Error | $\rho=0.5$ | Linear | 3 | 9021.77 | 23.8% | 25080.71 | 46.1% | 24793.8 | 45.5% |
| | | | 12 | 8401.76 | 23.3% | 24384.89 | 47.0% | 24798.8 | 47.4% |
| | | Quadratic | 3 | 8484.34 | 23.4% | 25430.00 | 45.2% | 24804.5 | 44.1% |
| | | | 12 | 8354.53 | 21.7% | 24416.41 | 46.2% | 24804.2 | 46.6% |
| | | Exponential | 3 | 9599.44 | 24.4% | 24391.34 | 47.0% | 24424.7 | 47.0% |
| | | | 12 | 8451.71 | 23.1% | 24366.85 | 43.4% | 24814.2 | 43.9% |
| | $\rho=0.9$ | Linear | 3 | 19450.12 | 12.9% | 435794.96 | 111.1% | 436790.1 | 111.1% |
| | | | 12 | 18688.94 | 12.6% | 428823.54 | 114.1% | 436707.6 | 115.7% |
| | | Quadratic | 3 | 18838.35 | 12.5% | 436151.64 | 108.0% | 436811.5 | 107.9% |
| | | | 12 | 18626.56 | 11.7% | 428848.47 | 112.0% | 436713.6 | 113.6% |
| | | Exponential | 3 | 20248.68 | 13.4% | 426208.49 | 113.5% | 427562.1 | 113.8% |
| | | | 12 | 18754.17 | 12.5% | 428802.77 | 105.1% | 436726.0 | 106.6% |

On the other hand, SemiparMF methodology is inferior to the two generalized additive models when the response is dominated by the high frequency covariate exhibited in Table 3. Specifically, this is true when the error correlation is moderate and functional form of the high frequency covariate is linear of quadratic.

Table 3. MSPE and MAPE for Uncorrelated Covariates, Varying Contribution of the Covariates to the Response

| | Error Correlation | Weights f(x)-z-w-e | No. of Subtime Points | SemiparMF | | GAM - Ordinary | | GAM - Summarized | |
|---|---|---|---|---|---|---|---|---|---|
| | | | | MSPE | MAPE | MSPE | MAPE | MSPE | MAPE |
| Without Misspecification Error | ρ=0.5 | 30-30-30-10 | 3 | 531.11 | 11.2% | 632.68 | 12.5% | 309.6 | 8.7% |
| | | | 12 | 195.72 | 7.0% | 351.57 | 9.4% | 315.7 | 8.9% |
| | | 50-20-20-10 | 3 | 1119.96 | 16.0% | 1181.85 | 16.7% | 281.9 | 8.3% |
| | | | 12 | 270.11 | 8.5% | 389.68 | 10.2% | 278.7 | 8.5% |
| | | 20-50-20-10 | 3 | 366.41 | 9.4% | 470.27 | 10.7% | 327.6 | 8.9% |
| | | | 12 | 217.08 | 7.2% | 347.62 | 9.2% | 335.2 | 9.0% |
| | | 20-20-50-10 | 3 | 328.52 | 9.0% | 503.47 | 11.5% | 360.5 | 9.7% |
| | | | 12 | 158.28 | 6.1% | 389.93 | 9.9% | 378.5 | 9.8% |
| | ρ=0.9 | 30-30-30-10 | 3 | 913.29 | 10.2% | 4742.84 | 26.1% | 4432.5 | 25.2% |
| | | | 12 | 344.92 | 6.0% | 4398.07 | 25.2% | 4437.5 | 25.4% |
| | | 50-20-20-10 | 3 | 1870.21 | 14.7% | 5291.34 | 27.5% | 4404.0 | 24.9% |
| | | | 12 | 474.90 | 7.5% | 4434.74 | 25.9% | 4399.5 | 25.8% |
| | | 20-50-20-10 | 3 | 640.27 | 8.6% | 4580.42 | 25.5% | 4450.5 | 25.1% |
| | | | 12 | 384.43 | 6.4% | 4394.77 | 24.8% | 4457.1 | 25.0% |
| | | 20-20-50-10 | 3 | 542.97 | 7.8% | 4614.47 | 26.2% | 4484.1 | 25.8% |
| | | | 12 | 278.71 | 5.0% | 4437.61 | 25.3% | 4501.2 | 25.5% |
| With Misspecification Error | ρ=0.5 | 30-30-30-10 | 3 | 8940.74 | 23.7% | 24902.90 | 45.9% | 24664.0 | 45.4% |
| | | | 12 | 8382.13 | 22.6% | 24370.87 | 45.5% | 24794.4 | 45.9% |
| | | 50-20-20-10 | 3 | 9877.22 | 25.6% | 25448.96 | 46.6% | 24635.3 | 45.2% |
| | | | 12 | 8481.93 | 23.2% | 24406.53 | 46.4% | 24756.2 | 46.7% |
| | | 20-50-20-10 | 3 | 8697.06 | 23.0% | 24740.99 | 45.4% | 24681.9 | 45.2% |
| | | | 12 | 8411.41 | 22.4% | 24369.49 | 44.8% | 24813.5 | 45.3% |
| | | 20-20-50-10 | 3 | 8625.71 | 23.3% | 24776.55 | 46.4% | 24716.2 | 46.3% |
| | | | 12 | 8335.20 | 22.5% | 24410.64 | 45.5% | 24858.8 | 46.0% |
| | ρ=0.9 | 30-30-30-10 | 3 | 19376.94 | 12.8% | 432648.78 | 110.6% | 433708.9 | 110.7% |
| | | | 12 | 18662.67 | 12.2% | 428809.15 | 110.3% | 436704.0 | 111.8% |
| | | 50-20-20-10 | 3 | 20673.65 | 14.2% | 433202.29 | 110.9% | 433670.7 | 110.7% |
| | | | 12 | 18795.55 | 12.6% | 428831.49 | 112.7% | 436655.6 | 114.2% |
| | | 20-50-20-10 | 3 | 19049.15 | 12.3% | 432490.36 | 109.7% | 433731.0 | 109.9% |
| | | | 12 | 18703.77 | 12.1% | 428810.43 | 108.5% | 436725.8 | 110.1% |
| | | 20-20-50-10 | 3 | 18949.79 | 12.5% | 432532.02 | 112.3% | 433774.5 | 112.5% |
| | | | 12 | 18597.56 | 12.1% | 428848.64 | 110.2% | 436777.6 | 111.7% |

The performance of the SemiparMF method and other two GAMs on different levels of sample size and length of time series is summarized in Table 4. Note that GAM with summarized values of $X_{it_k}$'s gave the lowest MSPE and MAPE specifically when the rate of its occurrence is low (K=3). This suggests that when the frequency of the covariate with higher temporal resolution is low, we can just opt to summarize its value into its mean as there will be not much information that will be lost in the estimation.

Table 4. MSPE and MAPE for Uncorrelated Covariates, Different Sample Size and Length of Time Series

| | Error Correlation | Sample Size and Length of Series | No. of Subtime Points | SemiparMF | | GAM - Ordinary | | GAM - Summarized | |
|---|---|---|---|---|---|---|---|---|---|
| | | | | MSPE | MAPE | MSPE | MAPE | MSPE | MAPE |
| Without Misspecification Error | ρ=0.5 | T=50; N=50 | 3 | 618.70 | 11.5% | 762.65 | 13.4% | 386.0 | 9.8% |
| | | | 12 | 231.92 | 7.4% | 437.37 | 10.5% | 393.3 | 9.9% |
| | | T=50; N=30 | 3 | 519.28 | 11.1% | 566.12 | 11.9% | 188.5 | 7.2% |
| | | | 12 | 164.36 | 6.7% | 240.02 | 8.2% | 195.6 | 7.4% |
| | | T=30; N=50 | 3 | 621.52 | 11.6% | 762.43 | 13.4% | 385.2 | 9.8% |
| | | | 12 | 234.61 | 7.5% | 431.71 | 10.4% | 392.2 | 9.9% |
| | ρ=0.9 | T=50; N=50 | 3 | 1034.72 | 9.9% | 6035.76 | 28.9% | 5668.3 | 28.0% |
| | | | 12 | 383.62 | 5.9% | 5653.12 | 28.1% | 5675.6 | 28.2% |
| | | T=50; N=30 | 3 | 819.02 | 10.6% | 2454.10 | 21.3% | 2085.5 | 19.8% |
| | | | 12 | 261.25 | 6.2% | 2092.37 | 19.9% | 2091.0 | 20.0% |
| | | T=30; N=50 | 3 | 1121.31 | 10.4% | 5931.94 | 28.8% | 5574.5 | 27.9% |
| | | | 12 | 467.35 | 6.5% | 5503.41 | 27.8% | 5579.8 | 28.1% |
| With Misspecification Error | ρ=0.5 | T=50; N=50 | 3 | 10990.96 | 24.9% | 31581.65 | 50.3% | 31278.4 | 49.8% |
| | | | 12 | 10360.03 | 23.9% | 31051.79 | 50.0% | 31406.4 | 50.3% |
| | | T=50; N=30 | 3 | 4687.49 | 21.1% | 11892.42 | 37.8% | 11566.2 | 37.0% |
| | | | 12 | 4096.47 | 19.4% | 11391.33 | 37.0% | 11610.7 | 37.4% |
| | | T=30; N=50 | 3 | 11427.09 | 25.6% | 31427.98 | 50.1% | 31178.4 | 49.7% |
| | | | 12 | 10751.49 | 24.6% | 30725.03 | 49.6% | 31400.0 | 50.3% |
| | ρ=0.9 | T=50; N=50 | 3 | 21070.03 | 12.0% | 555470.20 | 118.3% | 556365.8 | 118.3% |
| | | | 12 | 20262.81 | 11.4% | 552207.44 | 118.1% | 559394.4 | 119.3% |
| | | T=50; N=30 | 3 | 8505.27 | 11.6% | 199371.17 | 97.4% | 199905.0 | 97.4% |
| | | | 12 | 7723.46 | 10.5% | 196393.04 | 96.7% | 200841.3 | 98.3% |
| | | T=30; N=50 | 3 | 28961.85 | 15.2% | 543313.72 | 116.8% | 544893.0 | 117.0% |
| | | | 12 | 28083.40 | 14.8% | 537874.30 | 116.4% | 549911.4 | 118.3% |

It is further noted that superior predictive ability of the SemiparMF model is evident on scenarios with high (ρ =0.9) autocorrelation of error terms.

## 5.2 Covariates with weak autocorrelation (ρ=0.1)

Since the independent variables cannot always be assumed to be exogenous specially in a spatiotemporal system, scenarios when they have exhibit autocorrelation is included in the simulation settings.

Estimates in the presence of weak autocorrelation in the covariates are summarized in Table 5, 6 and 7. When autocorrelation is absent from the covariates, lower average MSPE and MAPE can be noted in GAM with summarized $X_{it_k}$'s in almost all cases regardless of the functional form of the more frequent covariate when $K = 3$ and error autocorrelation is moderate (**ρ**=0.5). When the functional form of the $X_{it_k}$'s is quadratic, better results are observed in SemiparMF over the other two GAMs in both cases of K as presented in Table 5.

Lower MSPE and MAPE is noted in SemiparMF model at any level of contribution of the covariates to the response except when autocorrelation of the error terms is moderate and K=3. Interestingly, errors when K=3 from SemiparMF algorithm was found to be twice as much as that of the summarized GAM approach particularly when the response is dominated by the covariate with higher temporal resolution.

Table 5. MSPE and MAPE for Weakly Autocorrelated Covariates (ρ=0.1) and Varying Functional Form of the Higher Frequency Covariate

| | Error Correlation | Functional Form | No. of Subtime Points | SemiparMF | | GAM - Ordinary | | GAM - Summarized | |
|---|---|---|---|---|---|---|---|---|---|
| | | | | MSPE | MAPE | MSPE | MAPE | MSPE | MAPE |
| Without Misspecification Error | ρ=0.5 | Linear | 3 | 565.64 | 10.9% | 692.51 | 12.2% | 316.2 | 8.2% |
| | | | 12 | 200.71 | 6.9% | 368.77 | 9.4% | 321.5 | 8.8% |
| | | Quadratic | 3 | 209.71 | 7.2% | 1128.71 | 14.7% | 327.0 | 7.9% |
| | | | 12 | 167.61 | 5.7% | 410.04 | 9.7% | 327.3 | 8.6% |
| | | Exponential | 3 | 1020.58 | 13.8% | 377.20 | 9.7% | 321.5 | 8.9% |
| | | | 12 | 248.74 | 7.5% | 350.98 | 8.4% | 337.3 | 8.2% |
| | ρ=0.9 | Linear | 3 | 927.60 | 9.9% | 4782.23 | 24.7% | 4420.6 | 23.7% |
| | | | 12 | 353.18 | 5.9% | 4400.29 | 24.8% | 4424.4 | 24.9% |
| | | Quadratic | 3 | 365.85 | 6.1% | 5217.90 | 24.8% | 4431.2 | 22.7% |
| | | | 12 | 301.69 | 4.9% | 4443.26 | 24.3% | 4430.0 | 24.3% |
| | | Exponential | 3 | 1605.71 | 12.6% | 4466.61 | 25.4% | 4425.9 | 25.3% |
| | | | 12 | 428.09 | 6.6% | 4380.02 | 23.0% | 4440.4 | 23.2% |
| With Misspecification Error | ρ=0.5 | Linear | 3 | 8965.17 | 22.6% | 25030.98 | 43.6% | 24728.2 | 43.1% |
| | | | 12 | 8381.34 | 22.3% | 24331.59 | 44.7% | 24724.1 | 45.1% |
| | | Quadratic | 3 | 8479.70 | 22.7% | 25473.61 | 42.6% | 24740.1 | 41.4% |
| | | | 12 | 8342.34 | 20.9% | 24374.79 | 43.7% | 24728.8 | 44.1% |
| | | Exponential | 3 | 9622.84 | 23.1% | 24705.65 | 45.7% | 24733.8 | 45.7% |
| | | | 12 | 8439.72 | 22.0% | 24307.36 | 41.7% | 24740.0 | 42.2% |
| | ρ=0.9 | Linear | 3 | 19383.67 | 12.2% | 433681.20 | 104.6% | 434890.7 | 104.7% |
| | | | 12 | 18698.50 | 12.1% | 426675.78 | 107.9% | 434747.5 | 109.5% |
| | | Quadratic | 3 | 18734.22 | 12.0% | 434123.02 | 101.0% | 434900.1 | 100.9% |
| | | | 12 | 18647.20 | 11.4% | 426725.43 | 105.5% | 434751.0 | 107.0% |
| | | Exponential | 3 | 20283.31 | 12.6% | 433358.77 | 110.8% | 434899.7 | 111.0% |
| | | | 12 | 18776.82 | 12.0% | 426633.02 | 100.8% | 434766.5 | 102.3% |

Table 6. MSPE and MAPE for Weakly Autocorrelated Covariates ($\rho$=0.1) and Different Contribution of the Covariates to the Response

| | Error Correlation | Weights f(x)-z-w-e | No. of Subtime Points | SemiparMF | | GAM - Ordinary | | GAM - Summarized | |
|---|---|---|---|---|---|---|---|---|---|
| | | | | MSPE | MAPE | MSPE | MAPE | MSPE | MAPE |
| Without Misspecification Error | $\rho$=0.5 | 30-30-30-10 | 3 | 538.36 | 10.4% | 663.52 | 11.9% | 311.0 | 8.2% |
| | | | 12 | 191.08 | 6.5% | 357.68 | 8.9% | 317.1 | 8.4% |
| | | 50-20-20-10 | 3 | 1165.91 | 14.9% | 1262.79 | 15.8% | 282.7 | 7.7% |
| | | | 12 | 267.97 | 7.8% | 401.71 | 9.7% | 279.1 | 8.0% |
| | | 20-50-20-10 | 3 | 363.80 | 8.9% | 485.46 | 10.4% | 329.1 | 8.6% |
| | | | 12 | 209.47 | 6.8% | 351.23 | 8.8% | 336.7 | 8.6% |
| | | 20-20-50-10 | 3 | 326.51 | 8.3% | 519.47 | 10.8% | 363.5 | 9.0% |
| | | | 12 | 154.24 | 5.6% | 395.78 | 9.3% | 381.9 | 9.1% |
| | $\rho$=0.9 | 30-30-30-10 | 3 | 888.44 | 9.5% | 4752.96 | 24.7% | 4415.4 | 23.8% |
| | | | 12 | 336.58 | 5.6% | 4388.71 | 23.9% | 4420.0 | 24.1% |
| | | 50-20-20-10 | 3 | 1827.76 | 13.5% | 5351.48 | 25.9% | 4386.0 | 23.4% |
| | | | 12 | 464.18 | 7.0% | 4436.21 | 24.4% | 4380.6 | 24.3% |
| | | 20-50-20-10 | 3 | 618.94 | 8.1% | 4573.86 | 24.5% | 4432.2 | 24.1% |
| | | | 12 | 369.74 | 6.0% | 4378.89 | 23.9% | 4438.3 | 24.1% |
| | | 20-20-50-10 | 3 | 530.41 | 7.2% | 4610.67 | 24.6% | 4470.0 | 24.2% |
| | | | 12 | 273.44 | 4.7% | 4427.63 | 23.8% | 4487.4 | 24.0% |
| With Misspecification Error | $\rho$=0.5 | 30-30-30-10 | 3 | 8930.46 | 22.6% | 25000.98 | 43.8% | 24723.3 | 43.3% |
| | | | 12 | 8368.16 | 21.7% | 24319.57 | 43.3% | 24719.0 | 43.7% |
| | | 50-20-20-10 | 3 | 9854.43 | 24.1% | 25607.16 | 44.2% | 24693.0 | 42.7% |
| | | | 12 | 8466.22 | 22.2% | 24369.98 | 44.0% | 24678.2 | 44.2% |
| | | 20-50-20-10 | 3 | 8685.30 | 22.3% | 24817.21 | 44.0% | 24740.5 | 43.8% |
| | | | 12 | 8392.75 | 21.7% | 24306.57 | 43.3% | 24737.8 | 43.8% |
| | | 20-20-50-10 | 3 | 8620.09 | 22.1% | 24854.97 | 44.0% | 24779.2 | 43.8% |
| | | | 12 | 8324.06 | 21.4% | 24355.52 | 42.9% | 24789.0 | 43.4% |
| | $\rho$=0.9 | 30-30-30-10 | 3 | 19336.04 | 12.1% | 433649.25 | 105.2% | 434885.8 | 105.3% |
| | | | 12 | 18682.00 | 11.8% | 426648.78 | 104.7% | 434744.1 | 106.1% |
| | | 50-20-20-10 | 3 | 20603.93 | 13.2% | 434255.42 | 104.6% | 434842.0 | 104.4% |
| | | | 12 | 18811.77 | 12.1% | 426743.51 | 106.1% | 434691.8 | 107.6% |
| | | 20-50-20-10 | 3 | 19010.40 | 11.9% | 433456.09 | 106.1% | 434898.1 | 106.3% |
| | | | 12 | 18716.82 | 11.8% | 426610.82 | 104.7% | 434750.0 | 106.2% |
| | | 20-20-50-10 | 3 | 18917.89 | 11.8% | 433523.24 | 105.9% | 434961.4 | 106.2% |
| | | | 12 | 18619.45 | 11.6% | 426709.20 | 103.6% | 434834.2 | 105.1% |

Table 7. MSPE and MAPE for Weakly Autocorrelated Covariates (**ρ**=0.1) and Different Sample Size and Length of Time Series

| | **Error Correlation** | **Functional Form** | **No. of Subtime Points** | **SemiparMF** | | **GAM - Ordinary** | | **GAM - Summarized** | |
|---|---|---|---|---|---|---|---|---|---|
| | | | | MSPE | MAPE | MSPE | MAPE | MSPE | MAPE |
| **Without Misspecification Error** | **ρ=0.5** | T=50; N=50 | 3 | 629.91 | 10.8% | 797.96 | 12.7% | 385.6 | 9.1% |
| | | | 12 | 226.85 | 6.9% | 441.67 | 9.9% | 392.6 | 9.3% |
| | | T=50; N=30 | 3 | 536.12 | 10.3% | 599.46 | 11.3% | 189.1 | 6.7% |
| | | | 12 | 159.69 | 6.2% | 244.99 | 7.8% | 196.3 | 6.9% |
| | | T=30; N=50 | 3 | 629.91 | 10.8% | 801.01 | 12.7% | 390.1 | 9.2% |
| | | | 12 | 230.52 | 7.0% | 443.14 | 9.9% | 397.1 | 9.4% |
| | **ρ=0.9** | T=50; N=50 | 3 | 1009.70 | 9.2% | 6033.53 | 27.4% | 5638.9 | 26.6% |
| | | | 12 | 374.66 | 5.6% | 5625.66 | 26.7% | 5643.8 | 26.8% |
| | | T=50; N=30 | 3 | 799.82 | 9.8% | 2468.21 | 20.1% | 2071.7 | 18.7% |
| | | | 12 | 249.35 | 5.7% | 2087.77 | 18.8% | 2079.5 | 18.9% |
| | | T=30; N=50 | 3 | 1089.65 | 9.6% | 5965.00 | 27.3% | 5567.0 | 26.5% |
| | | | 12 | 458.95 | 6.1% | 5510.14 | 26.5% | 5571.4 | 26.7% |
| **With Misspecification Error** | **ρ=0.5** | T=50; N=50 | 3 | 10985.69 | 23.8% | 31635.86 | 48.0% | 31300.9 | 47.5% |
| | | | 12 | 10350.10 | 23.0% | 30929.60 | 47.6% | 31301.1 | 47.9% |
| | | T=50; N=30 | 3 | 4670.23 | 20.0% | 11928.92 | 35.9% | 11573.1 | 35.1% |
| | | | 12 | 4080.60 | 18.5% | 11368.80 | 35.1% | 11575.8 | 35.5% |
| | | T=30; N=50 | 3 | 11411.79 | 24.5% | 31645.46 | 48.0% | 31328.0 | 47.6% |
| | | | 12 | 10732.69 | 23.7% | 30715.33 | 47.4% | 31316.0 | 47.9% |
| | **ρ=0.9** | T=50; N=50 | 3 | 21030.42 | 11.4% | 556092.70 | 112.6% | 557082.5 | 112.6% |
| | | | 12 | 20274.73 | 11.1% | 549377.72 | 112.3% | 556957.0 | 113.4% |
| | | T=50; N=30 | 3 | 8462.73 | 10.9% | 199525.59 | 92.1% | 200062.2 | 92.1% |
| | | | 12 | 7720.27 | 10.1% | 195332.55 | 91.3% | 200105.5 | 92.9% |
| | | T=30; N=50 | 3 | 28908.04 | 14.5% | 545544.72 | 111.7% | 547545.7 | 111.9% |
| | | | 12 | 28127.52 | 14.3% | 535323.97 | 110.7% | 547202.5 | 112.4% |

As seen in Table 7, note that when autocorrelation of the error terms is moderate (**ρ**=0.5), the SemiparMF model showed superior performance over the other two GAMs in cases when K=12 regardless of the sample size and length of series. On the other hand, the generalized additive model with summarized covariates observed thrice as frequent as the response (K=3) gave better estimates than the SemiparMF model. It is important to note that the SemiparMF model gave the lowest MSPE and MAPE in many cases when there are less units than the length of time series (N=30, T=50).

*5.3 Covariates with strong autocorrelation (ρ=0.8)*

Tables 8, 9 and 10 illustrate the performance of the three models when the covariates has strong autocorrelation (ρ=0.8). When the autocorrelation of the error terms is moderate (ρ=0.5), SemiparMF model generally showed superior ability over the generalized additive model when the covariate with higher temporal resolution is observed at a more frequent rate (K=12). On the other hand, when rate of occurrence of the more frequent covariate is low (K=3), summarized GAM yielded better results in most cases. This is consistent to the results from previous scenarios when there is no autocorrelation or weak autocorrelation in the covariates.

Cases when the series is near non-stationarity or error autocorrelation is strong (ρ=0.9), SemiparMF model has better predictive ability on all cases, with or without misspecification error.

However, note that when the functional form of the higher frequency covariate is quadratic, summarized GAM proved to be the better especially in cases with moderate autocorrelation of error terms (ρ=0.5).

Lowest MSPE and MAPE was calculated from summarized GAM on both cases of K=3 and K=12 regardless of weight combinations of the components as shown in Table 9. This was different from the usual result that SemiparMF model yield higher predictive abiity when K=12 over the other generalized additive models.

Table 8. MSPE and MAPE for Strongly Autocorrelated Covariates (ρ=0.8) and Varying Functional Form of the Higher Frequency Variable

| | Error Correlation | Functional Form | No. of Subtime Points | SemiparMF | | GAM - Ordinary | | GAM - Summarized | |
|---|---|---|---|---|---|---|---|---|---|
| | | | | MSPE | MAPE | MSPE | MAPE | MSPE | MAPE |
| Without Misspecification Error | ρ=0.5 | Linear | 3 | 506.50 | 4.5% | 1362.75 | 7.5% | 340.2 | 4.1% |
| | | | 12 | 166.82 | 3.2% | 476.78 | 5.6% | 344.4 | 4.9% |
| | | Quadratic | 3 | 4013.60 | 10.7% | 18041.19 | 13.1% | 479.4 | 2.4% |
| | | | 12 | 1462.83 | 5.6% | 2309.68 | 6.5% | 370.7 | 3.0% |
| | | Exponential | 3 | 6984.54 | 8.1% | 9896.87 | 17.2% | 993.4 | 5.5% |
| | | | 12 | 866.72 | 3.9% | 3727.74 | 9.1% | 667.2 | 4.2% |
| | ρ=0.9 | Linear | 3 | 653.29 | 4.1% | 5398.31 | 13.8% | 4419.5 | 12.7% |
| | | | 12 | 285.16 | 3.1% | 4304.08 | 14.3% | 4418.2 | 14.7% |
| | | Quadratic | 3 | 4252.98 | 9.4% | 22081.99 | 13.8% | 4560.5 | 7.2% |
| | | | 12 | 1659.39 | 5.2% | 6199.02 | 10.3% | 4443.6 | 9.2% |
| | | Exponential | 3 | 7233.37 | 7.5% | 13944.84 | 18.8% | 5079.8 | 12.3% |
| | | | 12 | 1035.91 | 3.7% | 7640.74 | 12.5% | 4739.3 | 10.4% |
| With Misspecification Error | ρ=0.5 | Linear | 3 | 8759.75 | 13.1% | 25469.87 | 24.8% | 24636.0 | 24.3% |
| | | | 12 | 8368.05 | 14.5% | 23556.62 | 26.9% | 24612.9 | 27.7% |
| | | Quadratic | 3 | 12263.73 | 14.7% | 42196.96 | 18.2% | 24781.2 | 14.2% |
| | | | 12 | 9650.91 | 11.4% | 25523.22 | 18.2% | 24638.0 | 18.0% |
| | | Exponential | 3 | 15217.04 | 10.3% | 34047.08 | 26.7% | 25303.0 | 22.9% |
| | | | 12 | 9050.96 | 10.1% | 26991.26 | 20.6% | 24937.8 | 19.9% |
| | ρ=0.9 | Linear | 3 | 19134.41 | 7.9% | 428550.39 | 62.3% | 432150.6 | 62.7% |
| | | | 12 | 18855.90 | 8.9% | 403875.53 | 66.7% | 431482.4 | 70.0% |
| | | Quadratic | 3 | 23318.46 | 9.0% | 445336.82 | 40.9% | 432324.3 | 40.3% |
| | | | 12 | 20426.50 | 7.6% | 406461.36 | 46.9% | 431497.3 | 49.0% |
| | | Exponential | 3 | 26682.26 | 7.2% | 437247.65 | 59.5% | 432900.0 | 59.0% |
| | | | 12 | 19691.03 | 6.9% | 408143.66 | 50.8% | 431789.4 | 52.9% |

The SemiparMF also also yield higher predictive ability in cases when there is misspecification error and when there is strong autocorrelation in the covariates. Note that MSPE and MAPE are significantly lower than that of the other GAMs specially in cases with high autocorrelation in the error terms.

Lowest MSPE and MAPE was calculated from summarized GAM when there is moderate autocorrelation of error terms for both balanced and unbalanced data sets as shown in Table 10. Conversely, SemiparMF model yield better results in most cases with strong autocorrelation of error terms (ρ=0.9). This further shows the strength of

SemiparMF model in spatiotemporal data with covariates measured at a higher frequency (K>3) than the response.

SemiparMF model is still able to chanracterize the relationship between the response and the covariates even in the presence of misspecification error.

Table 9. MSPE and MAPE for Strongly Autocorrelated Covariates ($\rho$=0.8) and Different Contribution of the Covariates to the Response

| P | Error Correlation | Functional Form | No. of Subtime Points | SemiparMF | | GAM - Ordinary | | GAM - Summarized | |
|---|---|---|---|---|---|---|---|---|---|
| | | | | MSPE | MAPE | MSPE | MAPE | MSPE | MAPE |
| Without Misspecification Error | $\rho$=0.5 | 30-30-30-10 | 3 | 3299.17 | 7.9% | 8406.59 | 12.8% | 553.5 | 3.8% |
| | | | 12 | 724.99 | 4.2% | 1895.58 | 7.0% | 430.1 | 3.7% |
| | | 50-20-20-10 | 3 | 8953.75 | 9.0% | 22724.14 | 14.5% | 904.9 | 3.1% |
| | | | 12 | 1801.88 | 4.5% | 4631.09 | 7.3% | 543.2 | 2.8% |
| | | 20-50-20-10 | 3 | 1568.70 | 7.2% | 4015.08 | 11.7% | 528.6 | 4.9% |
| | | | 12 | 428.08 | 4.4% | 1130.37 | 7.2% | 486.3 | 5.0% |
| | | 20-20-50-10 | 3 | 1517.90 | 7.1% | 3921.93 | 11.5% | 430.3 | 4.4% |
| | | | 12 | 373.56 | 4.0% | 1028.57 | 6.8% | 383.5 | 4.5% |
| | $\rho$=0.9 | 30-30-30-10 | 3 | 3516.95 | 7.1% | 12448.22 | 15.1% | 4635.6 | 10.2% |
| | | | 12 | 887.86 | 4.0% | 5772.19 | 11.9% | 4503.0 | 10.9% |
| | | 50-20-20-10 | 3 | 9204.62 | 8.2% | 26768.24 | 15.2% | 4987.9 | 7.7% |
| | | | 12 | 1999.15 | 4.3% | 8554.25 | 10.3% | 4613.3 | 8.2% |
| | | 20-50-20-10 | 3 | 1760.03 | 6.4% | 8051.56 | 15.7% | 4606.8 | 12.5% |
| | | | 12 | 574.85 | 4.0% | 4979.75 | 13.6% | 4557.1 | 13.3% |
| | | 20-20-50-10 | 3 | 1704.60 | 6.3% | 7965.48 | 15.7% | 4516.1 | 12.5% |
| | | | 12 | 512.08 | 3.7% | 4885.60 | 13.6% | 4461.4 | 13.3% |
| With Misspecification Error | $\rho$=0.5 | 30-30-30-10 | 3 | 11541.18 | 12.4% | 32544.74 | 22.4% | 24855.9 | 19.6% |
| | | | 12 | 8915.61 | 11.6% | 25081.90 | 21.1% | 24698.9 | 21.1% |
| | | 50-20-20-10 | 3 | 17200.37 | 11.2% | 46887.82 | 19.6% | 25205.9 | 14.8% |
| | | | 12 | 9982.58 | 9.5% | 27912.37 | 16.9% | 24806.1 | 16.1% |
| | | 20-50-20-10 | 3 | 9814.29 | 13.6% | 28132.11 | 25.3% | 24825.3 | 23.7% |
| | | | 12 | 8623.49 | 13.4% | 24260.93 | 24.7% | 24751.6 | 25.1% |
| | | 20-20-50-10 | 3 | 9764.86 | 13.7% | 28053.87 | 25.5% | 24739.8 | 23.8% |
| | | | 12 | 8571.55 | 13.4% | 24172.94 | 24.8% | 24661.8 | 25.2% |
| | $\rho$=0.9 | 30-30-30-10 | 3 | 22455.96 | 7.9% | 435681.15 | 52.5% | 432407.3 | 52.2% |
| | | | 12 | 19531.75 | 7.6% | 405889.21 | 53.2% | 431559.2 | 55.6% |
| | | 50-20-20-10 | 3 | 28958.02 | 7.7% | 450074.50 | 42.1% | 432760.9 | 41.2% |
| | | | 12 | 20817.05 | 6.7% | 409169.58 | 42.8% | 431639.9 | 44.6% |
| | | 20-50-20-10 | 3 | 20416.15 | 8.2% | 431215.57 | 61.0% | 432337.2 | 61.1% |
| | | | 12 | 19175.95 | 8.4% | 404800.70 | 61.4% | 431587.2 | 64.3% |
| | | 20-20-50-10 | 3 | 20350.05 | 8.3% | 431208.59 | 61.4% | 432327.7 | 61.5% |
| | | | 12 | 19106.50 | 8.4% | 404781.25 | 61.7% | 431572.5 | 64.6% |

Table 10. MSPE and MAPE for Strongly Autocorrelated Covariates ($\rho$=0.8) and Different Sample Size and Length of Time Series

| | Error Correlation | Functional Form | No. of Subtime Points | SemiparMF | | GAM - Ordinary | | GAM - Summarized | |
|---|---|---|---|---|---|---|---|---|---|
| | | | | MSPE | MAPE | MSPE | MAPE | MSPE | MAPE |
| Without Misspecification Error | $\rho$=0.5 | T=50; N=50 | 3 | 3826.04 | 7.8% | 9802.68 | 12.6% | 666.8 | 4.3% |
| | | | 12 | 849.91 | 4.3% | 2266.26 | 7.3% | 527.8 | 4.3% |
| | | T=50; N=30 | 3 | 3757.73 | 7.7% | 9519.04 | 12.5% | 465.9 | 3.5% |
| | | | 12 | 778.75 | 4.1% | 2027.53 | 6.7% | 329.4 | 3.4% |
| | | T=30; N=50 | 3 | 3920.86 | 7.8% | 9979.09 | 12.7% | 680.3 | 4.3% |
| | | | 12 | 867.72 | 4.4% | 2220.41 | 7.2% | 525.0 | 4.3% |
| | $\rho$=0.9 | T=50; N=50 | 3 | 4051.07 | 6.8% | 14968.63 | 16.2% | 5909.9 | 12.1% |
| | | | 12 | 1012.01 | 3.9% | 7271.87 | 13.6% | 5756.1 | 12.9% |
| | | T=50; N=30 | 3 | 3851.22 | 7.1% | 11321.00 | 13.9% | 2339.1 | 8.1% |
| | | | 12 | 852.34 | 3.8% | 3828.86 | 10.1% | 2199.2 | 8.6% |
| | | T=30; N=50 | 3 | 4237.35 | 7.1% | 15135.50 | 16.2% | 5810.9 | 12.0% |
| | | | 12 | 1116.11 | 4.2% | 7043.12 | 13.4% | 5645.7 | 12.8% |
| With Misspecification Error | $\rho$=0.5 | T=50; N=50 | 3 | 14008.51 | 13.4% | 40406.50 | 25.0% | 31513.9 | 22.8% |
| | | | 12 | 10992.21 | 12.9% | 31872.45 | 24.3% | 31337.8 | 24.3% |
| | | T=50; N=30 | 3 | 7717.64 | 11.1% | 20682.98 | 19.4% | 11778.1 | 15.9% |
| | | | 12 | 4677.22 | 9.8% | 12866.88 | 17.4% | 11639.5 | 17.0% |
| | | T=30; N=50 | 3 | 14514.37 | 13.7% | 40624.42 | 25.2% | 31428.2 | 22.8% |
| | | | 12 | 11400.49 | 13.2% | 31331.77 | 24.0% | 31211.5 | 24.3% |
| | $\rho$=0.9 | T=50; N=50 | 3 | 24580.05 | 7.6% | 558603.19 | 59.0% | 555468.3 | 58.9% |
| | | | 12 | 21146.58 | 7.4% | 525778.39 | 60.0% | 554114.9 | 62.4% |
| | | T=50; N=30 | 3 | 11859.94 | 7.3% | 205651.69 | 45.2% | 198864.1 | 44.7% |
| | | | 12 | 8455.19 | 6.5% | 188574.71 | 45.8% | 198410.0 | 47.6% |
| | | T=30; N=50 | 3 | 32695.14 | 9.1% | 546879.98 | 58.6% | 543042.5 | 58.4% |
| | | | 12 | 29371.66 | 9.4% | 504127.46 | 58.6% | 542244.3 | 61.8% |

## 6. Illustration

The empirical feasibility of the SemiparMF algorithm is tested in the estimation of a spatiotemporal model for corn yield using remotely-sensed data measured at a higher frequency. The data used here is the quarterly corn production data from Philippine Statistical Authority for the period 2003-2015. The study area covered the provinces in Luzon island, comprising 7 regions or a total of 38 provinces. Figure 1 shows the map of the study area and the provinces within it (in dark shades).

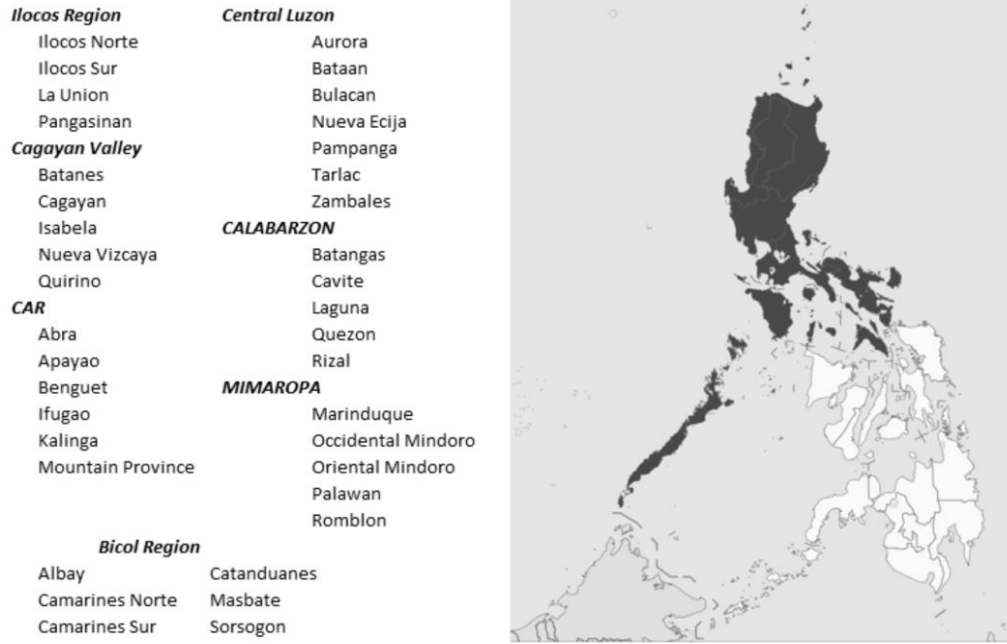

Figure 1. Map of Luzon Island and its provinces

*The Data*

*Yield*

Corn production data was downloaded from the Philippine Statistics Authority website: (http://countrystat.psa.gov.ph/). Yield is defined as total production per area harvested with corn. The dependent variable is yield in metric tons per hectare.

*Vegetation Indices*

Moderate Resolution Imaging Spectroradiometer (MODIS) is an instrument that operates both in Aqua and Terra spacecraft. It has a viewing swath width of 2,330 km and views the entire surface of the Earth every one to two days. Its detectors measure 36 spectral bands and it acquires data at three spatial resolutions: 250-m, 500-m, and 1,000-m. Vegetation index data used in this study were extracted from MODIS product MYD13A3. This product is a monthly composite of MODIS data at 1-km spatial resolution. These datasets were downloaded from Land Processes Distributed Active

Archive Center (LP DAAC) (https://lpdaac.usgs.gov/dataset_discovery/modis/modis_products_table/myd13a3). Raw normalized vegetation indices extracted from these files were then postprocessed and smoothed using TIMESAT tool.

To extract the value for each province, the smoothed values are then rasterized using the *raster* package in R. Monthly vegetation time series for each province are constructed by taking the average of the indeces measured inside the area of each province per month. Vegetation indices are multiplied by 0.0001 for proper scaling.

*Precipitation Estimates*

Since we don't have complete rain gauge sensors to measure the amount of rain that fall in each province, we used precipitation estimates from the operation Tropical Rainfall Measuring Mission (TRMM) Multisatellite Precipitation Analysis (TMPA). TRMM is a satellite launched in 1997 that provides critical precipitation estimates in the tropical and subtropical regions of the Earth. In April 2015, the instruments on this mission has been turned off and is currently being transitioned to the Global Precipitation Measurement (GPM) mission. Rainfall estimates for the Philippines that is used in this study were extracted from the 3B42 version 7 of the TRMM product downloaded from NASA Goddard Earth Sciences (GES) Data and Information Services Center (DISC). (http://disc.sci.gsfc.nasa.gov/SSW/#keywords=TRMM_3B42_daily%207) Estimates from this product has 0.25º by 0.25º grid cell spatial resolution with coverage from 50N–50S. Accuracy of rainfall estimates from TRMM and other satellite products was studied by Bagtasa et al (2016).

Raw extracts from the TRMM files are first divided by 25 because each grid size is approximately 25km by 25km. Since there are multiple grid boxes that cover a province, we take the average of the rainfall estimates from these grid boxes. These daily average rainfall estimates (in mm/km2) are then summed up per quarter to finally create the quarterly rainfall estimates per province.

*Fertilizer Applied*

We added another variable into the model that will capture intervention to enhance corn production. The variable included in the model is the average inorganic fertilizer applied in corn farms. The data was obtained from agricultural statistics of the Philippine Statistics Authority. This time series is reported annually at regional level. Since the available data is only up to 2014, we append our 2015 estimate to the data by fitting an AR(1) model for each time series per region. This will also be used as a neighboring system that is defined at regional grouping of the provinces.

*Estimated Model*

The SemiparMF model was slightly modified in the estimation of corn yield using remotely-sensed data. Since rainfall pattern varies across the Luzon island, we allow the parameter $\beta_i$ to vary across the provinces. The temporal parameter $\rho_i$ is also allowed to vary among the provinces. The semiparametric spatiotemporal model for corn yield is expressed in Equation (8)

$$y_{it} = \sum_{k=1}^{3} f(X_{it_k}) + \beta_i Z_{it} + \gamma_i W_{it} + \varepsilon_{it} \qquad (8)$$

$$\varepsilon_{it} = \rho_i \varepsilon_{it-1} + a_{it}, |\rho_i| < 1 \quad a_{it} \sim IID(0, \sigma_a^2), i = 1, .., n, t = 1, .. T$$

where

$Y_{it}$ = corn yield, in metric tons per hectare in province *i* at quarter *t*

$X_{it_k}$ = average vegetation index in province *i* at month *k* of quarter *t*, k=1,2,3

$f(\cdot)$ = a continuous function in $X_{it_k}$

$Z_{it}$ = total accumulated rainfall (precipitation index) in mm/km2 in province $i$ at quarter $t$

$W_{it}$ = quantity of inorganic fertilizer applied to corn in the region where province $i$ belongs at quarter $t$

$\varepsilon_{it}$ = error terms for province $i$ at time $t$

*Prediction of Corn Yield*

The predictive performance of the SemiparMF model for corn yield evaluated by computing the MSPE and MAPE. The model is also compared to generalized additive models used in the simulation studies. The errors are summarized in Table 11.

Table 11. MSPE and MAPE of Corn Yield Model

|  | Prediction Error | | | | | MSPE | MAPE |
|---|---|---|---|---|---|---|---|
|  | Min | 1st Quartile | Median | 3rd Quartile | Max | | |
| **SemiparMF** | -3.17 | -0.45 | 0.03 | 0.43 | 2.69 | 0.54 | 28% |
| **GAM – Ordinary** | -4.01 | -0.93 | -0.01 | 0.9 | 4.01 | 1.54 | 58% |
| **GAM - Summarized** | -4.05 | -0.97 | -0.04 | 0.96 | 4.23 | 1.59 | 59% |

After 3 iterations, the backfitting algorithm in SemiparMF model has already converged. Comparing this hybrid model to two other additive models, we can observe superior predictive ability of the model in terms of both MSPE and MAPE. The SemiparMF model takes advantage over GAM in terms of the unnecessary aggregation of the high frequency data to complement those of the low frequency data.

## 7. Conclusion

The SemiparMF model is capable of producing better model with good predictive ability over ordinary generalized additive models. Specifically, superior performance on various cases when the rate of occurrence of the more frequent covariate is high (K=12) regardless of its functional form and autocorrelation of the covariates. This supports the value of the methodology in optimizing the use of the unaggregated level of the higher frequency covariate in explaining the variability of the response.

The simulation scenarios also showed that the SemiparMF estimated with a hybrid algorithm demonstrated better predictive ability than the other two GAMs when the autocorrelation of the error terms is high. The SemiparMF model also proved to be useful when there is misspecification error as illustrated by almost twice as lower MSPE and MAPE than the generalized additive models.